\documentclass[prb,
numerical,
sd,
amsmath,
amssymb,
notitlepage,
superscriptaddress
]{revtex4-1}

\usepackage{amssymb} 
\usepackage{amsthm}
\usepackage{graphicx,subfigure}
\usepackage{dcolumn}
\usepackage{bm}
\usepackage{cancel}
\usepackage{float}
\usepackage{mathtools}
\usepackage{epstopdf}
\usepackage{hyperref}\hypersetup{colorlinks=true, linkcolor=blue, urlcolor=blue, citecolor=blue}
\usepackage{url}
\usepackage{array}
\usepackage{xcolor}
\usepackage{tabularx}
\allowdisplaybreaks

\def\nuebar{\bar{\nu}_e}

\begin{document}

\title{AGM2015: Antineutrino Global Map 2015}

\author{S.M. Usman}
\affiliation{Exploratory Science and Technology Branch, National Geospatial-Intelligence Agency, Springfield, VA, 22150, USA}
\email*{Shawn.Usman@nga.mil}

\author{G.R. Jocher}
\affiliation{Ultralytics LLC, Arlington, VA, 22203, USA}

\author{S.T. Dye}
\affiliation{Department of Physics and Astronomy, University of Hawaii, Honolulu, HI, 96822, USA}
\affiliation{Department of Natural Sciences, Hawaii Pacific University, Kaneohe, HI, 96744, USA}

\author{W.F. McDonough}
\affiliation{Department of Geology, University of Maryland, College Park, MD, 20742, USA}

\author{J.G. Learned}
\affiliation{Department of Physics and Astronomy, University of Hawaii, Honolulu, HI, 96822, USA}

\maketitle

\textbf{Every second greater than $10^{25}$ antineutrinos radiate to space from Earth, shining like a faint antineutrino star. Underground antineutrino detectors have revealed the rapidly decaying fission products inside nuclear reactors, verified the long-lived radioactivity inside our planet, and informed sensitive experiments for probing fundamental physics. Mapping the anisotropic antineutrino flux and energy spectrum advance geoscience by defining the amount and distribution of radioactive power within Earth while critically evaluating competing compositional models of the planet. We present the Antineutrino Global Map 2015 (AGM2015), an experimentally informed model of Earth's surface antineutrino flux over the 0 to 11 MeV energy spectrum, along with an assessment of systematic errors. The open source AGM2015 provides fundamental predictions for experiments, assists in strategic detector placement to determine neutrino mass hierarchy, and aids in identifying undeclared nuclear reactors. We use cosmochemically and seismologically informed models of the radiogenic lithosphere/mantle combined with the estimated antineutrino flux, as measured by KamLAND and Borexino, to determine the Earth's total antineutrino luminosity at  $3.4^{+2.3}_{-2.2} \times 10^{25} \nuebar$/s. We find a dominant flux of geo-neutrinos, predict sub-equal crust and mantle contributions, with $\sim$1\% of the total flux from man-made nuclear reactors.}

\begin{figure}[h]
\includegraphics[width=1\linewidth]{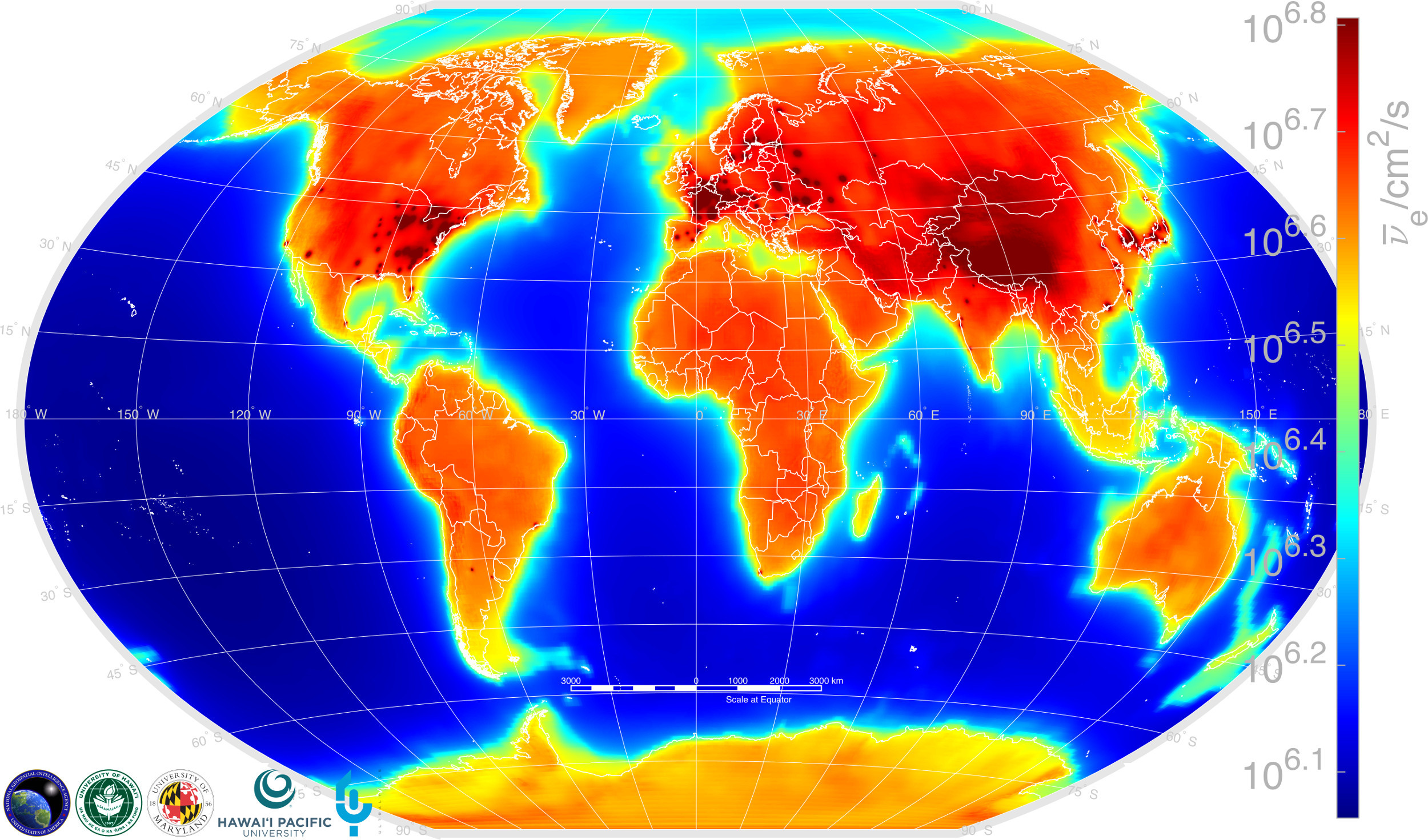}
\caption{AGM2015: A worldwide $\nuebar$ flux map combining geoneutrinos from natural $^{238}$U and $^{232}$Th decay in the Earth's crust and mantle as well as manmade reactor-$\nuebar$ emitted by power reactors worldwide. Flux units are $\nuebar/\mathrm{cm}^2/\mathrm{s}$ at the Earth's surface. Map includes $\nuebar$ of all energies. Figure created with MATLAB \cite{matlab}.}
\label{worldall}
\end{figure}

\section*{Introduction}

The neutrino was proposed by Wolfgang Pauli in 1930 to explain the continuous energy spectrum of nuclear beta rays. By Pauli's hypothesis the missing energy was carried off by a lamentably ``undetectable" particle. Enrico Fermi succeeded in formulating a theory for calculating neutrino emission in tandem with a beta ray \cite{fer34}. Detecting Pauli's particle required exposing many targets to an intense neutrino source. While working on the Manhattan Project in the early 1940s Fermi succeeded in producing a self-sustaining nuclear chain reaction, which by his theory was recognized to copiously produce antineutrinos. Antineutrino detection projects were staged near nuclear reactors the following decade. In 1955, Raymond Davis, Jr. found that reactor antineutrinos did not transmute chlorine to argon by the reaction: $^{37}$Cl ($\bar{\nu_e}$, e$^-$) $^{37}$Ar \cite{davis55}. This result permitted the existence of Pauli's particle only if neutrinos are distinct from antineutrinos. Davis later used the chlorine reaction to detect solar neutrinos using 100,000 gallons of dry-cleaning fluid deep in the Homestake Gold Mine. Reactor antineutrinos were ultimately detected in 1956 by Clyde Cowan and Fred Reines by recording the transmutation of a free proton by the reaction $^{1}$H ($\bar{\nu_e}$, e$^+$) $^{1}$n \cite{rei56, cow56}. This detection confirmed the existence of the neutrino and marked the advent of experimental neutrino physics. 

Almost 60 years later neutrino research remains an active and fruitful pursuit in the fields of particle physics, astrophysics, and cosmology. In addition to nuclear reactors and the Sun, detected neutrino sources include particle accelerators \cite{AGS}, the atmosphere \cite{reines,kolar}, core-collapse supernovae \cite{kamII,IMB3,baksan,Hax13}, the Earth \cite{kam13, bor13}, and most recently the cosmos \cite{icecube}. We now know that neutrinos and antineutrinos have ``flavor" associations with each of the charged leptons (e, $\mu$, $\tau$) and these associations govern their interactions. Neutrino flavors are linear combinations of neutrino mass eigenstates ($\nu_1$, $\nu_2$, and $\nu_3$). This quantum mechanical phenomenon, known as neutrino oscillation, changes the probability of detecting a neutrino in a given flavor state as a function of energy and distance. Neutrino flavor oscillations along with their low cross section provide a glimpse into some of the most obscured astrophysical phenomena in the universe and most recently the otherwise inaccessible interior of our planet. Antineutrinos emanating from the interior of our planet constrain geochemical models of Earth's current radiogenic interior. Antineutrino observations of the modern Earth's interior coupled with cosmochemical analysis of chronditic meteorites from the early solar system allow scientists to model the geochemical evolution of the Earth across geologic time.


\begin{figure*}[!htbp]
\includegraphics[width=\linewidth]{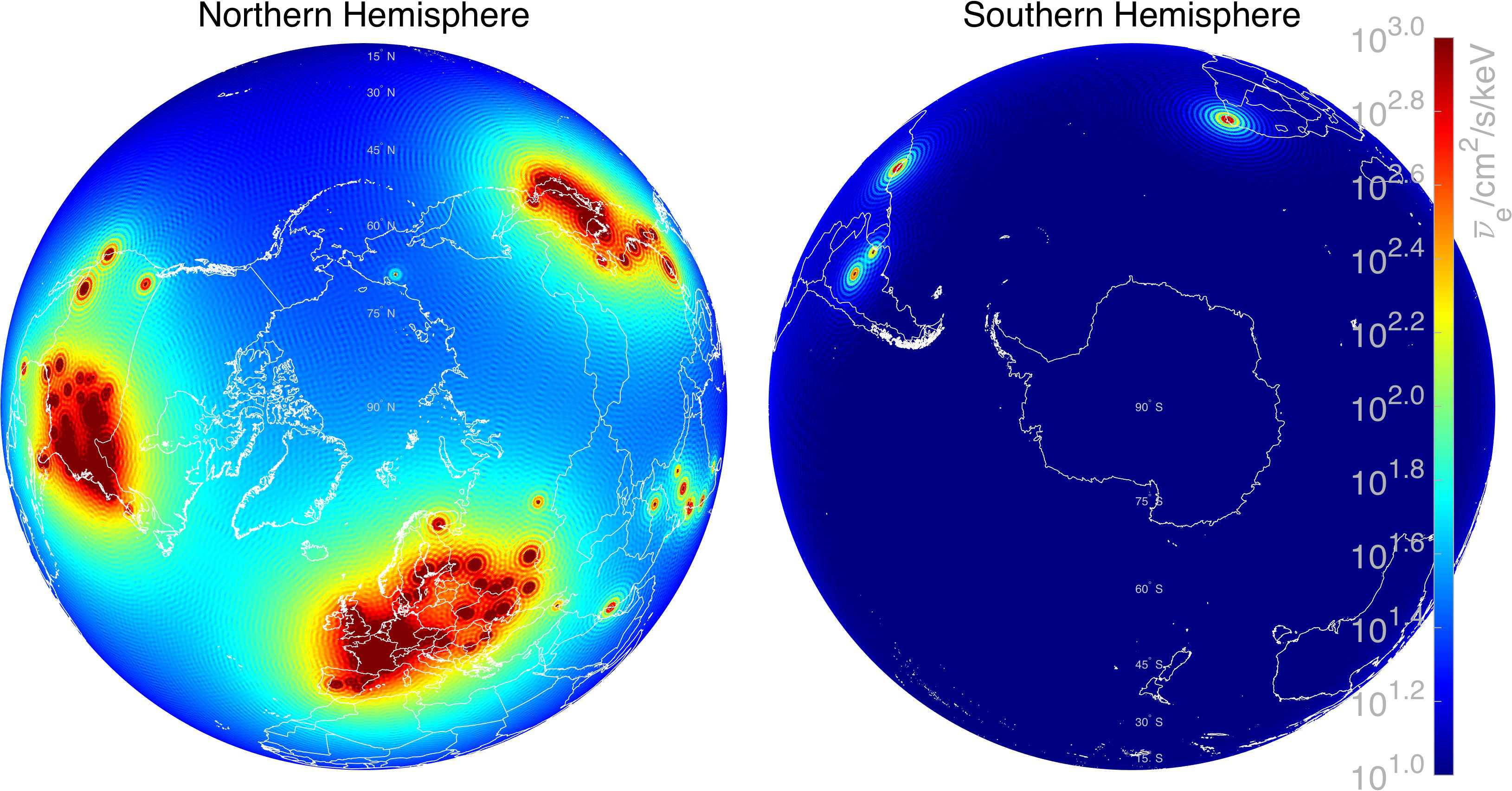}
\caption{AGM2015 reactor-$\nuebar$ flux in the 3.00-3.01MeV energy bin (in logspace color). Flux units are $\nuebar/\mathrm{cm}^2/\mathrm{s}$ at the Earth's surface.  Note the visible $\theta_{12}$ oscillations at $\sim$100km wavelength. Figure created with MATLAB \cite{matlab}.}
\label{fluxmap3MeVreactor}
\end{figure*}

Recently, the blossoming field of neutrino geoscience, first proposed by Eder \cite{ede66}, has become a reality with 130 observed geoneutrino interactions \cite{kam13, bor13} confirming Kobayashi's view of the Earth being a ``neutrino star" \cite{Kob91}. These measurements have constrained the radiogenic heating of the Earth along with characterizing the distribution of U and Th in the crust and mantle. The development of next generation antineutrino detectors equipped with fast timing ($\thicksim$50ps) multichannel plates \cite{grabas13} coupled with Gd/Li doped scintillator will allow for the imaging of antineutrino interactions. The imaging and subsequent reconstruction of antineutrino interactions produce directionality metrics. Directionality information can be leveraged for novel geological investigations such as the geo-neutrinographic imaging of felsic magma chambers beneath volcanos \cite{tan14}. These exciting geophysical capabilities have significant overlap with the non-proliferation community where remote monitoring of antineutrinos emanating from nuclear reactors is being seriously considered \cite{ber10}.  

\begin{figure*}[!htbp]
\includegraphics[width=\linewidth]{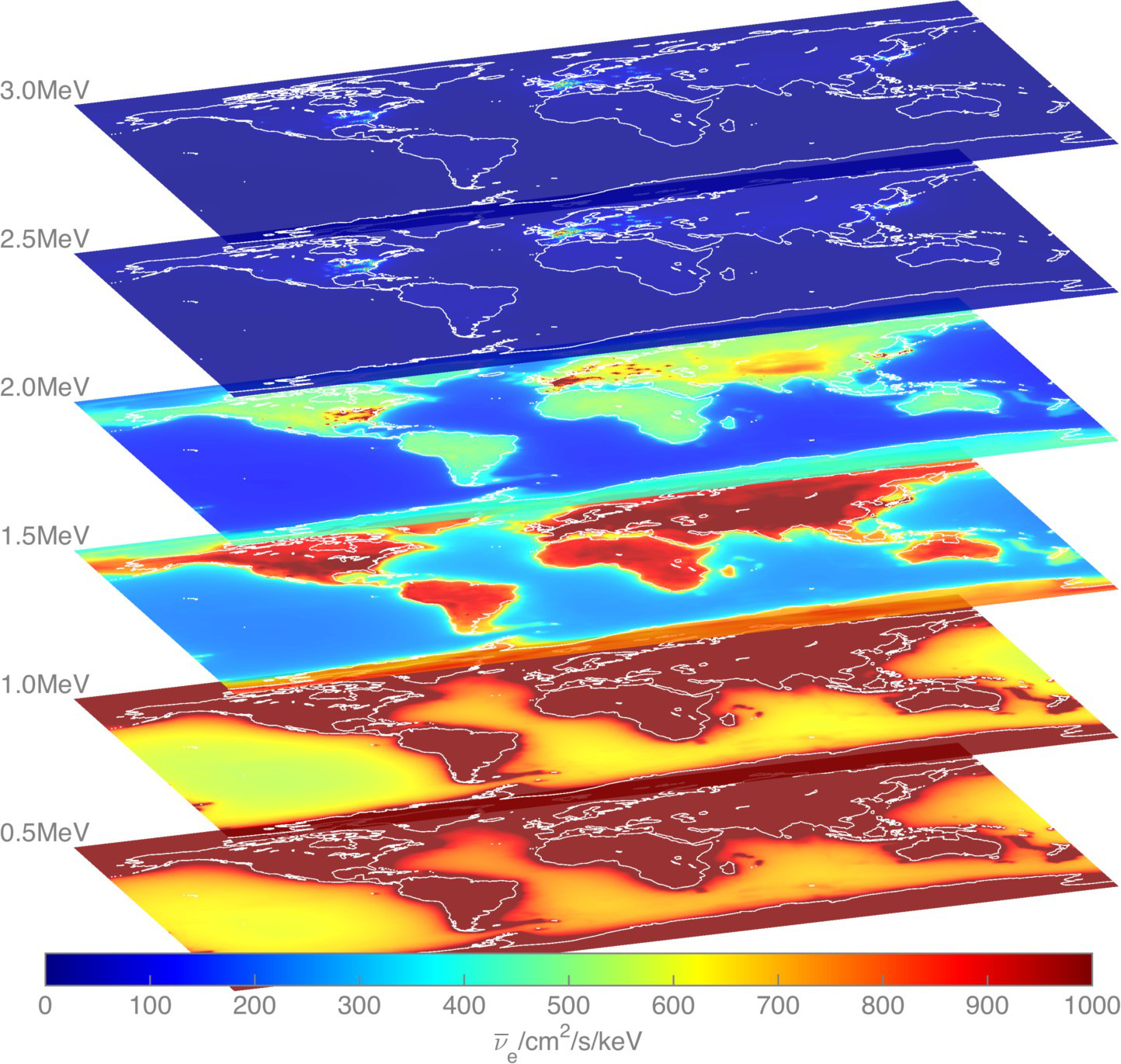}
\caption{AGM2015 $\nuebar$ flux ($\nuebar$/cm$^2$/s/keV) displayed at 6 select energy bins out of the 1100 total AGM2015 energy bins, which uniformly span the 0MeV - 11MeV $\nuebar$ energy range. Each energy bin is 10keV wide. In conjunction with 720 longitude bins and 360 latitude bins, the highest resolution AGM2015 map is a $360\times 720\times 1100$ 3D matrix comprising $\sim$300 million elements total. Figure created with MATLAB \cite{matlab}.}
\label{AGMlayers}
\end{figure*}

Antineutrino Global Map 2015 (AGM2015) shown in Figure \ref{worldall} merges geophysical models of the Earth into a unified energy dependent map of $\nuebar$ flux, both natural and manmade, at any point on the Earth's surface. We provide the resultant flux maps freely to the general public in a variety of formats at \href{http://www.ultralytics.com/agm2015}{www.ultralytics.com/agm2015}. AGM2015 aims to provide an opensource infrastructure to easily incorporate future neutrino observations that enhance our understanding of Earth's antineutrino flux and its impact on the geosciences. In this study we first describe the particle physics parameters used in propagating antineutrino oscillations across the planet's surface as shown in Figure \ref{fluxmap3MeVreactor}. A detailed description of the incorporation of anthropogenic and geophysical neutrino energy spectrum from 0-11 MeV is presented which allows for the four-dimensional generation (latitude, longitude, flux, and energy) of the antineutrino map as shown in separate energy bins in Figure \ref{AGMlayers}. A vertically stratified model of the Earth's density, shown in Figure \ref{world8layers}, based on seismological derived density models are combined with a cosmochemical elemental abundances to determine the geological signal of antineutrinos. This signal is then constrained by geo-neutrino measurements from KamLAND and Borexino and first order uncertainties associated with AGM map are then presented. 

\section*{Neutrino Oscillations}

AGM2015 incorporates the known 3-flavor oscillation behavior of antineutrinos. This starts with the standard 3-flavor Pontecorvo Maki Nakagawa Sakata (PMNS) matrix $U$:

\begin{equation}
\begin{split}
U &=
\left(\begin{array}{ccc} U_{e1} & U_{e2} & U_{e3}\\ U_{\mu1} & U_{\mu2} & U_{\mu3}\\ U_{\tau1} & U_{\tau2} & U_{\tau3} \end{array}\right)\\
  &=
\left(\begin{array}{ccc} 1 & 0 & 0\\ 0 & {{c_{2}}}_{3} & {{s_{2}}}_{3}\\ 0 & - {{s_{2}}}_{3} & {{c_{2}}}_{3} \end{array}\right)
\left(\begin{array}{ccc} {{c_{1}}}_{3} & 0 & {{s_{1}}}_{3}e^{-i\delta}\\ 0 & 1 & 0\\ - {{s_{1}}}_{3}e^{i\delta} & 0 & {{c_{1}}}_{3} \end{array}\right)
\left(\begin{array}{ccc} {{c_{1}}}_{2} & {{s_{1}}}_{2} & 0\\ - {{s_{1}}}_{2} & {{c_{1}}}_{2} & 0\\ 0 & 0 & 1 \end{array}\right)
\left(\begin{array}{ccc} 1 & 0 & 0\\ 0 & e^{i\alpha_1/2} & 0\\ 0 & 0 & e^{i\alpha_2/2} \end{array}\right)\\
\end{split}
\end{equation}

\noindent where $c_{ij}=\cos\left({\theta_{ij}}\right)$ and $s_{ij}=\sin\left({\theta_{ij}}\right)$, and $\theta_{ij}$ denotes the neutrino oscillation angle from flavor $i$ to flavor $j$ in radians. In this paper we assume 

\begin{equation}
\begin{split}
\theta_{12}	& = 0.587 ^{+0.019}_{-0.017} \\
\theta_{13}	& = 0.152 ^{+0.007}_{-0.008}
\end{split}
\end{equation}

 \noindent per a global fit by Fogli {\it et al.}\cite{fogli2012} in the case of $\theta_{12}$, and by measurements at the Day Bay experiment \cite{daya} in the case of $\theta_{13}$. Phase factors $\alpha_1$ and $\alpha_2$ are nonzero only if neutrinos are Majorana particles (i.e. if neutrinos and antineutrinos are their own antiparticles), and have no influence on the oscillation survival probabilities, only on the rate of possible neutrino-less double beta decay. We assume $\alpha_1=\alpha_2=0$ in this work. We likewise assume phase factor $\delta=0$, though this assumption may change in the future if evidence is found to support neutrino oscillations violating charge parity (CP) symmetry.

The probability of a neutrino originally of flavor $\alpha$ later being observed as flavor $\beta$ is:

\begin{equation}
\begin{split}
P_{\alpha \rightarrow \beta}  & = {\left| \left\langle\nu_\beta | \nu_\alpha (t) \right\rangle \right|}^2 \\
& = {\left| \sum\limits_{i}^{}U_{\alpha i}^{*}U_{\beta i} e^{-i m_{i}^{2} L/2E} \right|}^2 \\
& = \delta_{\alpha \beta}
-4\sum\limits_{j}\sum\limits_{i>j} \Re \left(U_{\alpha i}^{*}U_{\beta i}U_{\alpha j}U_{\beta j}^{*} \right)\sin^2\left(\frac{\Delta m_{ij}^{2}L}{4E}\right)
+2\sum\limits_{j}\sum\limits_{i>j} \Im\left(U_{\alpha i}^{*}U_{\beta i}U_{\alpha j}U_{\beta j}^{*} \right)\sin^2\left(\frac{\Delta m_{ij}^{2}L}{4E}\right) \\
& \approx 1
-4\sum\limits_{j}\sum\limits_{i>j} U_{\alpha i}^{}U_{\beta i}U_{\alpha j}U_{\beta j}^{} \sin^2\left(1.267 \frac{\Delta m_{ij}^{2}}{\mathrm{eV^2}}\frac{L}{\mathrm{km}}\frac{\mathrm{GeV}}{E}\right)
\label{baseosceqn}
\end{split}
\end{equation}

\noindent
where $E$ is the neutrino energy in GeV, $L$ is the distance from its source the neutrino has traveled in km, and the delta-mass term $\Delta m_{ij}^{2} = m_{i}^{2} - m_{j}^{2}$, in eV$^2$. The last approximation assumes no charge parity (CP) violation ($\delta=0$), causing the imaginary terms to fall out. The * symbol denotes a complex conjugate, and $U_{ij}$ denotes the element of the PMNS matrix $U$ occupying the $i^{th}$ row and $j^{th}$ column. Equation \ref{baseosceqn} can be employed to determine the `survival probability' of a $\bar{\nu}_e$ of energy $E$ GeV later being observed as the same flavor a distance $L$ km from its source. In particular, the $P_{e \rightarrow e}$ survival probability of most interest to this paper can be expressed as Equation \ref{osceqn1}:

\begin{equation}
\begin{split}
P_{e \rightarrow e} = 1 - &4\, c^2_{12}\, c^4_{13}\, s^2_{12}\, {\sin\!^2\left(\frac{1.27 L \Delta m^2_{21}}{E}\right)}
 - 4\, c^2_{12}\, c^2_{13}\, s^2_{13}\, {\sin\!^2\left(\frac{1.27 L \Delta m^2_{31}}{E}\right)}
 - 4\, c^2_{13}\, s^2_{12}\, s^2_{13}\, {\sin\!^2\left(\frac{1.27 L \Delta m^2_{32}}{E}\right)}
\end{split}
\label{osceqn1}
\end{equation}

\noindent
For simplicity we ignore the Mikheyev-Smirnov-Wolfenstein (MSW) effect \cite{Wolfenstein} on neutrinos as they travel through the Earth. We use the neutrino mixing angles and mass constants from Fogli \textit{et al.} 2012\cite{fogli2012} and Daya Bay 2014\cite{daya} to evaluate Equation \ref{osceqn1} for all source-observer ranges and energies used in AGM, giving us the survival probability of seeing each source from each point in the map at each energy level. This is not a trivial task, requiring $> 1\times 10^{15}$ evaluations of Equation \ref{osceqn1} for a full AGM2015 rendering. This is broken down into $\sim 1 \times 10^6$ point sources, $\sim 1 \times 10^6$ locations on the map at which the flux is evaluated, and $\sim 1 \times 10^3$ energy bins spanning the $0< E_{\nuebar}<11$MeV energy range as shown in Figure \ref{AGMlayers}. Equation \ref{osceqn1} can best be visualized in Figure \ref{fluxmap3MeVreactor}, which shows the $\theta_{12}$ ripples in the 3MeV worldwide reactor-$\nuebar$ flux.

\section*{Reactor Antineutrinos} 

Reactor $\nuebar$ experiments have proven the viability of unobtrusive reactor monitoring and continue to contribute important information on neutrino properties including the possibility of additional light ``sterile" neutrino flavors \cite{aba12}.  We use the International Atomic Energy Agency's (IAEA) Power Reactor Information System (PRIS)\cite{PRIS2014} to identify and locate 435 known man-made reactor cores in operation at the time of this writing. PRIS categorizes reactors into four states:

\begin{itemize} \itemsep1pt \parskip0pt \parsep0pt
\item Operational
\item Under Construction
\item Temporary Shutdown
\item Permanent Shutdown
\end{itemize}

AGM2015 includes all ``Operational" or ``Temporary Shutdown" reactors, including many reactors in Japan affected by the Fukushima-Daiichi disaster, which are classified as ``Temporary Shutdown" rather than ``Permanent Shutdown." PRIS shows 435 Operational and Temporary Shutdown reactor cores distributed among 193 sites, with 870 GW$_{\mathrm{th}}$ total output after load factor considerations, and 72 reactor cores among 42 sites (total 156 GW$_{\mathrm{th}}$ at 100\% load factor) currently Under Construction. The PRIS database reports thermal capacity directly, which is typically about three times electrical capacity (most reactors are about 30\% efficient in converting heat into electricity).  Historical ``load factors" of each core are used to convert the total thermal capacity to projected current and future thermal power output. Load factors account for the down-time related to maintenance and other outages, allowing for AGM2015 to be a reliable estimate of worldwide antineutrino flux in its year of release. Typical PRIS load factors range from 70\% to 90\%. The 3MeV AGM2015 reactor-$\nuebar$ flux map due to these 435 Operational and Temporary Shutdown reactor cores is shown in Figure \ref{fluxmap3MeVreactor}.

\begin{figure*}[!htbp]
\includegraphics[width=\linewidth]{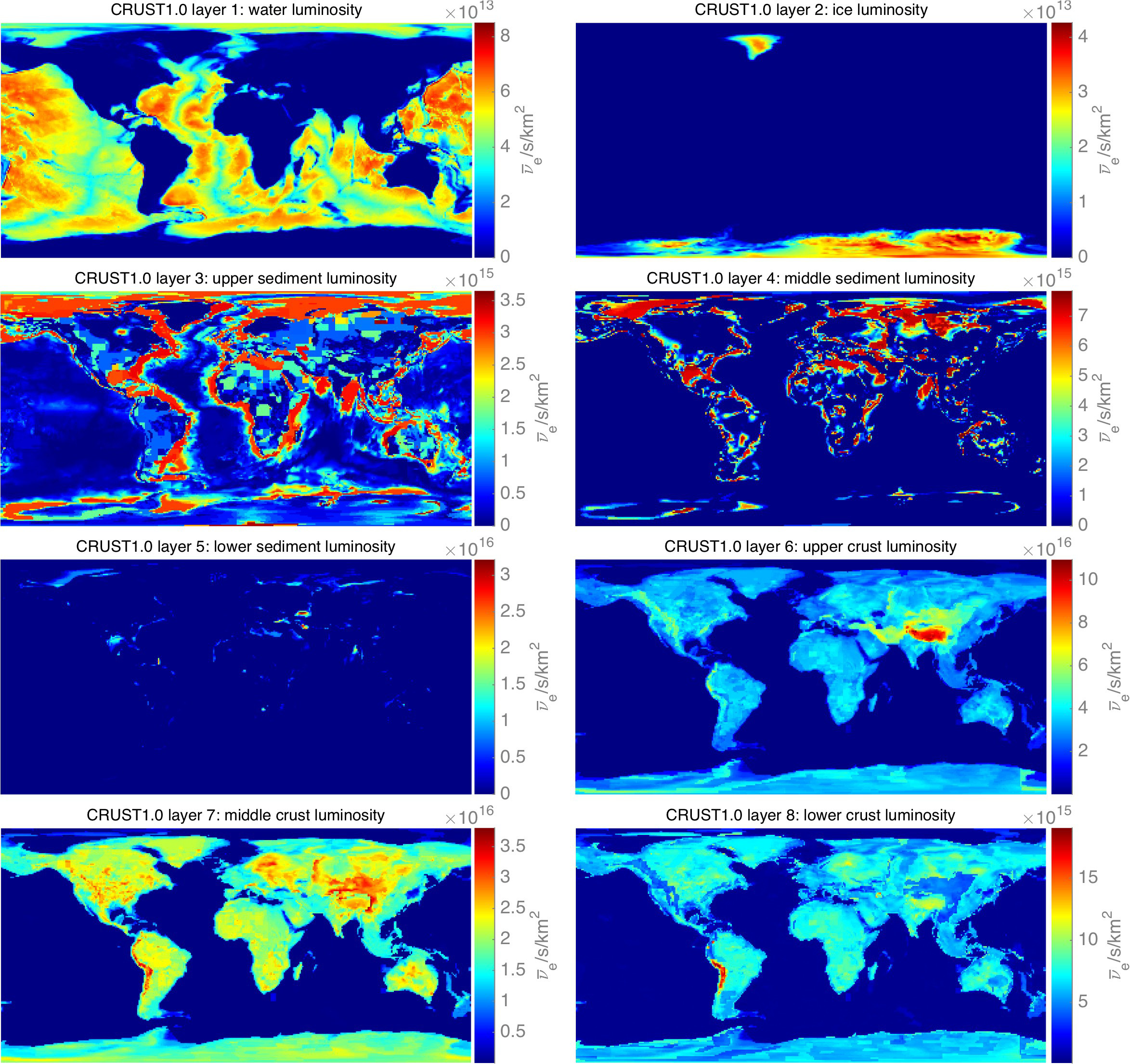}
\caption{AGM2015 $\nuebar$ luminosity per km$^2$ per CRUST1.0 layer. Each CRUST1.0 layer is composed of 180$\times$360 $1^\circ$ tiles, each with a defined thickness (ranging from 0-78 km) and density (ranging from 0.9-3.4 g/$\mathrm{cm}^3$). Note the layers have different colorbar scales. Figure created with MATLAB \cite{matlab}.}
\label{world8layers}
\end{figure*}

The reactor-$\nuebar$ energy spectrum assumes the shape of an exponential falloff in energy, with most reactor-$\nuebar$ released at the lowest energies. To obtain an $\nuebar /s / \mathrm{GW_{th}}$ reactor scaling we assume:

\begin{itemize} \itemsep1pt \parskip0pt \parsep0pt
\item The mean energy released per fission $E_f$ is around 205 MeV\cite{Reboulleau2010}.
\item The reactor thermal power $P_\mathrm{th}$ is related to the number of fissions per second $N_f = 6.24\times 10^{21} s^{-1} P_\mathrm{th}\mathrm{[GW]}/ E_f \mathrm{[MeV]}$\cite{Reboulleau2010}.
\item 6 $\nuebar$ created per fission\cite{ber10}
\item 2 $\nuebar$ created per fission on average above the inverse beta decay (IBD) energy threshold of $E_{\nu}\geq 1.8$ MeV\cite{ber10}. 
\end{itemize}

These assumptions yield $1.83 \times 10^{20}$ $\nuebar/\mathrm{s}/\mathrm{GW_{th}}$ (of all energies) emitted from a reactor, with $0.61 \times 10^{20}$ $\nuebar/\mathrm{s}/\mathrm{GW_{th}}$ emitted above the 1.8 MeV IBD detection threshold.  This is a mean value representative of a typical pressurized water reactor at the beginning of its fuel cycle \cite{ber10}. We find the summed worldwide reactor power output to be 870GW$_{\mathrm{th}}$, and the resultant $\nuebar$ luminosity to be $1.6 \times 10^{23} \nuebar / s$ and 0.04TW. Jocher {\it et al.}\cite{jocher2013} is recommended for a deeper discussion of reactor-$\nuebar$ detection via IBD detectors.

\section*{Geoneutrinos }

Observations from geology, geophysics, geochemistry, and meteoritics allow for a range of non-unique solutions for the composition of the Earth. The relative proportion of Fe, O, Mg, and Si in chondritic meteorites individually varies by $\sim15\%$ each and reflects spatial and temporal differences in where these rocks formed in the early solar nebula. Likewise, refractory elements have 25\% variation in their relative abundance, which translates into a factor of two in absolute concentration difference of these elements. Even greater enrichment factors of these elements occur when the volatile inventory (e.g., H$_{2}$O, CO$_{2}$, N$_{2}$) is mostly lost, as during terrestrial planet assembly. Finally, because the Earth's core is taken to have negligible amounts of Th, U and K\cite{mcdonough03,hitoshi2013,lay2008}, due to their limited solubility in core-forming metallic liquids, this becomes another 50\% enrichment factor in the radiogenic elements in the silicate Earth. Consequently, compositional models predict between 10 and 30 ng/g U (and Th/U = 3.9, the chondritic ratio) for the silicate Earth. Given the planetary ratio of Th/U and K/U ($1.4 \times 10^{4}$) \cite{are09}, and the absolute U content of the silicate Earth, its heat production for a 10 ng/g U model roughly corresponds to a surface heat flow of 10 TW and likewise 30 ng/g U to $\sim 30$ TW. Estimates of the Earth's radiogenic heat production thus vary from low power models (10-15 TW of power from K, Th, and U), through medium power models (17-22 TW), and to high power models ($>$ 25 TW) \cite{dye15}. Accordingly, detecting the Earth's flux of geoneutrinos can provide crucial data to test competing theories of the bulk Earth.

Two observatories, one in Japan (KamLAND) and one in Italy (Borexino), are making ongoing measurements of the surface flux of geoneutrinos at energies above the IBD threshold energy $E_{\nu}\geq 1.8$ MeV. At Japan the flux measurement is $(3.4\pm0.8)\times10^6$ cm$^{-2}$ s$^{-1}$ \cite{kam13}, while at Italy the flux measurement is $(4.3\pm1.3)\times10^6$ cm$^{-2}$ s$^{-1}$ \cite{bor13}. Note that it is sometimes convenient  to express geoneutrino flux as a rate of recorded interactions in a perfect detector with a given exposure using the Terrestrial Neutrino Unit (TNU) \cite{mantovani_2004}, however in this work we focus on simple $\nuebar$ flux ($\nuebar / \mathrm{cm}^2 / \mathrm{s}$) and luminosity ($\nuebar / \mathrm{s}$).

\begin{table}
\fontsize{9}{11} \selectfont \centering
\setlength\extrarowheight{3pt}
\setlength{\tabcolsep}{8pt}
\newcolumntype{n}{>{$}c<{$}} 
\begin{tabular}{l n}
$\Delta m^2_{12}$\cite{fogli2012}								&7.54^{+0.21}_{-0.18}  \times 10^{-5} \mathrm{eV}^2	\\
$\Delta m^2_{13}$\cite{daya}									&2.59^{+0.19}_{-0.20}  \times 10^{-3}\mathrm{eV}^2	\\ \hline
$\sin^2\theta_{12}$\cite{fogli2012}								&0.307^{+0.017}_{-0.016} 			\\
$\sin^2\theta_{13}$\cite{daya}									&0.02303^{+0.00210}_{-0.00235} 		\\ \hline
steady-state $\nuebar$ survival fraction\cite{fogli2012,daya}	&0.549\pm 0.012  					\\
Reactor $\nuebar$/fission\cite{Reboulleau2010,daya}				&6.00\pm 0.18  \hspace{1mm} 		\\
Reactor Energy[MeV]/fission\cite{Reboulleau2010,kopeikin2004}	&205 \pm 1 \hspace{1mm}				\\
Reactor $\nuebar$/fission$>$1.8 MeV \cite{Reboulleau2010,daya} 	&2.100 \pm 0.013 \hspace{1mm}		\\
\end{tabular}
\caption{AGM2015 reactor-$\nuebar$ parameters and $\nuebar$ oscillation parameters. $\theta_{12}$ oscillation parameters from Fogli {\it et al.}\cite{fogli2012}, and $\theta_{13}$ oscillation parameters from results of the Daya Bay\cite{daya} $\nuebar$ experiment.}
\label{table:reactor_elements}
\end{table}

\begin{table*}
\fontsize{9}{11} \selectfont \centering
\setlength\extrarowheight{3pt}
\setlength{\tabcolsep}{8pt}
\newcolumntype{n}{>{$}c<{$}} 
\begin{tabular}{c l n n n}
 &										& \mathrm{U}\hspace{1mm} (10^{-6})	& \mathrm{Th}\hspace{1mm} (10^{-6})  & \mathrm{K}\hspace{1mm} (10^{-2})	\\ \cline{3-5}
 & H$_{2}$O \cite{mantovani_2004}   	& 0.0032					& 0							&  0.04	 			\\
 & Sediment\cite{hua13}    				& 1.73\pm 0.09	 			& 8.10\pm 0.59				&  1.83\pm 0.12		\\ \cline{2-5}
 & Upper Crust\cite{hua13}    			& 2.7\pm 0.6				& 10.5\pm 1.0				&  2.32\pm 0.19	 	\\

 CC & Middle Crust\cite{hua13}    		& 0.97^{+0.58}_{-0.36}	 	& 4.86^{+4.30}_{-2.25}		&  1.52^{+0.81}_{-0.52} 	\\ 
 & Lower Crust\cite{hua13}    			& 0.16^{+0.14}_{-0.07}	 	& 0.96^{+1.18}_{-0.51}		&  0.65^{+0.34}_{-0.22}	 \\ \cline{2-5}

 OC & Crust\cite{hua13}    				& 0.07\pm 0.02		 		& 0.21\pm 0.06				&  0.07\pm 0.02		\\ \cline{2-5}
 & LM\cite{hua13}    					& 0.03^{+0.05}_{-0.02}		& 0.15^{+0.28}_{-0.10}		&  0.03^{+0.04}_{-0.02}	 \\
 & Mantle\cite{hua13,kam13, bor13,are09}   				& 0.011\pm 0.009 			& 0.022\pm 0.040			&  0.015\pm 0.013		\\  \noalign{\vskip 2mm} 
 &										& ^{238}\mathrm{U }			& ^{232}\mathrm{Th }  		& ^{40}\mathrm{K }	\\ \cline{3-5}
 & Isotope abundance  					& 0.99275	 				& 1.0						&  0.000117	 		\\  
 & Lifetime (Gyr)\cite{blu95,beg01}   	& 6.4460 	 				& 20.212					&  1.8005	 			\\ 
 & Multiplicity ($\nuebar$/decay)   	& 6	 						& 4							&  0.893	 			\\  
 & Mass (amu)   						& 238.05 					& 232.04					&  39.964	 	\\  

\end{tabular}
\caption{AGM2015 distribution and properties of U, Th, and K, which are the main emitters of electron antineutrinos. Abundances in the various stratified crustal layers are shown at the top of the table, including Oceanic Crust (OC), Continental Crust (CC), and Lithospheric Mantle (LM). Relevant isotopic properties are presented at the bottom of the table. Note the abundances are unit-less fractions. Uncertainties shown in this table derived from Huang {\it et al.} \cite{hua13}, Arevalo {\it et al.} \cite{are09}, Gando {\it et al.} \cite{kam13}, and Bellini {\it et al.} \cite{bor13}. }
\label{table:geo_elements}
\end{table*}

AGM models the Earth as a 3D point cloud consisting of roughly 1 million points. National Oceanic and Atmospheric Administration (NOAA) Earth TOPOgraphical 1 (ETOPO1) ``ice" data \cite{ETOPO1} is used to provide worldwide elevations with respect to the World Geodetic System 84 (WGS84) ellipsoid. Zero-tide ocean surface corrections to the WGS84 ellipsoid were obtained from the National Geospatial-Intelligence Agency (NGA) Earth Gravitational Model 2008 \cite{EGM2008} (EGM2008) for modeling the ocean surface elevations around the world. Underneath these surface elevations we model 8 separate crust layers using CRUST 1.0\cite{crust1.0}, shown in Figure \ref{world8layers}, as well as a 9th adjoining layer per Huang \textit{et al.}\cite{hua13} which reaches down to the spherical mantle, creating a seamless earth model. Certain crust tiles which are too large (about 200 km across at the equator) to be adequately modeled as point sources are instead modeled as collections of smaller tiles using numerical integration, which recursively subdivides large tiles into progressively smaller sub-tiles until the contribution of each is less than 0.001 TNU. 

Geoneutrino flux is produced from the decay of naturally occurring radioisotopes in the mantle and crust:   $^{238}$U, $^{232}$Th, $^{235}$U, $^{40}$K, $^{87}$Rb, $^{113}$Cd, $^{115}$In, $^{138}$La, $^{176}$Lu, and $^{187}$Re \cite{Enomoto_Sanshiro}. However, we only consider $^{238}$U and $^{232}$Th in our flux maps as all other elements' energy spectrum is considerably below the IBD energy threshold of $E_{\nu}\geq 1.8$ MeV. All abundances for the crust and mantle can be seen in Table \ref{table:geo_elements}. As shown in Table \ref{table:luminosities}, K is the largest contributor to $\nuebar$ luminosity but its energy is below the IBD threshold. All elements other than U, Th, and K have a negligible contribution to the Earth's $\nuebar$ luminosity. 

Successful detection of $\nuebar$ below 1.8 MeV remains elusive; if successful the incorporation of the remaining radioisotopes would be beneficial to future versions of AGM. The Earth's core was assumed to have no significant contribution to the $\nuebar$ flux due to limiting evidence for a georeactor \cite{dye06} and no appreciable amount of  $^{238}$U, $^{232}$Th, or $^{40}$K isotopes\cite{mcdonough03}. While certain core models support upper limits of K content at the $\sim$100 ppm level\cite{hitoshi2013}, which would be sufficient for up to $\sim$1-2TW of radiogenic heating in the present day, ``constraints on K content are very weak"\cite{lay2008}, and in the absence of stronger evidence we've chosen to assume a K-free core.

Mantle abundances were derived from empirical geo-neutrino measurements at KamLAND \cite{kam13} and Borexino \cite{bor13}. We deconstructed the reported geo-neutrino flux from each observation into separate contributions from U and Th according to a Th/U ratio of 3.9. From each of these, we subtracted the predicted crust flux contributions \cite{hua13} at each observatory, averaging the asymmetric non-gaussian errors, to arrive at estimates of the mantle contributions. We then combined the estimates of the mantle U flux and the mantle Th flux contributions from each observation in a weighted average. The resulting best estimates for the mantle U and Th flux contributions were finally converted to homogeneously distributed mantle abundances using the spherically symmetric density profile of the Preliminary Reference Earth Model (PREM) \cite{PREM} along with a corresponding correction to account for neutrino oscillations. Corresponding values for K were found by applying a K/U ratio of $13,800 \pm 1300$ \cite{are09}. The resulting AGM2015 U, Th and K mantle abundances are presented in Table \ref{table:geo_elements}. The main sources of uncertainty in these estimates are the observational errors in the flux measurements and limited knowledge of the subtracted crust fluxes. A detailed description of the methods and relevant conversion factors used here are presented in Dye \cite{dye15}.

\begin{figure*}[!htbp]
\includegraphics[width=\linewidth]{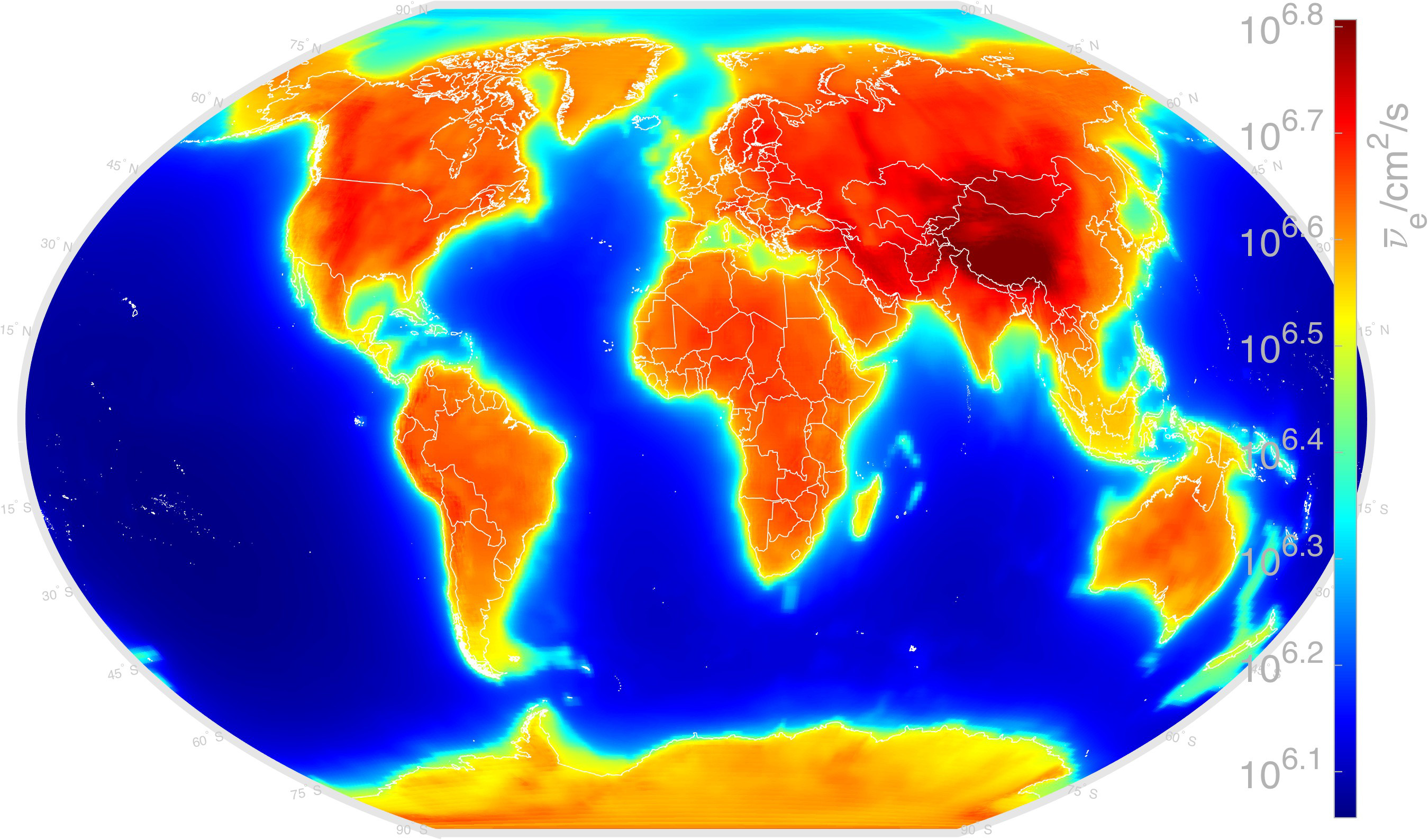}
\caption{AGM2015 geoneutrino flux due to $^{238}$U and $^{232}$Th decay in the Earth's crust and mantle. Flux units are $\nuebar/\mathrm{cm}^2/\mathrm{s}$ at the Earth's surface. Map includes $\nuebar$ of all energies. Figure created with MATLAB \cite{matlab}.}
\label{crustplusgeostacked}
\end{figure*}

AGM2015 neutrino luminosities are for total numbers of neutrinos. Although almost all are originally emitted as electron antineutrinos, on average only $\sim$0.55 of the total remain so due to neutrino oscillations. We calculate the total Earth $\nuebar$ luminosity to be $3.4^{+2.3}_{-2.2} \times10^{25}$ $ \nuebar$ s$^{-1}$. A detailed breakdown of $^{238}$U, $^{232}$Th, and $^{40}$K geoneutrino luminosity from the lithosphere and mantle can be seen in Table \ref{table:luminosities} (for all energies), as well as in Table \ref{table:luminosities2} for $E_{\nu}\geq 1.8$ MeV. Figure \ref{crustplusgeostacked} shows the combined AGM2015 crust+mantle $\nuebar$ flux.

\begin{table*}
\fontsize{9}{11} \selectfont \centering
\setlength\extrarowheight{5pt}
\setlength{\tabcolsep}{8pt}
\newcolumntype{n}{>{$}c<{$}} 
\begin{tabular}{ l n n n n n n}
							& 							& 							& L (10^{25} \nuebar/s)		&							&									\\ \cline{2-6} 
							& ^{238}\mathrm{U}			& ^{232}\mathrm{Th}			& ^{40}\mathrm{K}			& \mathrm{Reactors}			& \sum_{\mathrm{U,Th,K,Reactors}}	\\ \hline
Crust   						& 0.21^{+0.065}_{-0.050}	& 0.19^{+0.073}_{-0.042} 	&  0.91 ^{+0.22}_{-0.17}	& -							& 1.3^{+0.36}_{-0.26}	 			\\
Mantle    						& 0.32\pm{0.28}	 	& 0.14\pm{0.26}		&  1.6\pm{1.4}	 	& - 						& 2.1\pm{1.9}					\\ \hline
$\sum_\mathrm{Crust,Mantle}$    & 0.53^{+0.34}_{-0.33}		& 0.33^{+0.33}_{-0.30}  	&  2.5^{+1.6}_{-1.6}		& 0.016^{+0.001}_{-0.001} 	& 3.4^{+2.3}_{-2.2} 				\\
\end{tabular}
\caption{Contribution of geoneutrino luminosities $L$ in AGM2015 for $^{238}$U, $^{232}$Th, and $^{40}$K $\nuebar$ emitted by the Earth. The reactor-$\nuebar$ luminosity is $1.6 \pm 0.1 \times 10^{23} \nuebar / s$ and 40 GW for the world's 435 reactor cores, which together output 870GW$_\mathrm{th}$\cite{PRIS2014}. Uncertainties shown in this table derived from Huang {\it et al.} \cite{hua13}, Arevalo {\it et al.} \cite{are09}, Gando {\it et al.} \cite{kam13}, and Bellini {\it et al.} \cite{bor13}. }
\label{table:luminosities}
\end{table*}

A comparison of panels in Figure \ref{world8layers} reveals a more luminous upper crustal signature for the Tibetan plateau, Himalayan and Zagros Mountains, and the North American Cordillera. Whereas the lower crust of the central Andes Mountains is more luminous compared to the aforementioned bright regions of the upper crust which  on average are less luminous.  One potential explanation for contrasting observations might be repeated upper crustal thrusting in the Himalayas and like regions, while magmatic additions to the lower crust of the Andean arc  could account for their relative brightness.  These findings deserve further investigation to understand better their origin and distribution, particularly from seismic and geological perspectives.

\begin{table*}
\fontsize{9}{11} \selectfont \centering
\setlength\extrarowheight{5pt}
\setlength{\tabcolsep}{4pt}
\newcolumntype{n}{>{$}c<{$}} 
\begin{tabular}{ l n n n n n n}
								& 							& L (10^{25}\nuebar/s)>1.8\mathrm{MeV}	& 				&							&									\\ \cline{2-6} 
								& ^{238}\mathrm{U}			& ^{232}\mathrm{Th}				& ^{40}\mathrm{K}		& \mathrm{Reactors}			& \sum_{\mathrm{U,Th,K,Reactors}}	\\ \hline
fraction $>$1.8MeV 					& 0.068						& 0.042							&  0.0	 				& 0.35						& -	 								\\ \hline
Crust   							& 0.014^{+0.0044}_{-0.0034}	& 0.0082^{+0.0031}_{-0.0018} 	&  0.0	 				& -							& 0.022^{+0.0075}_{-0.0052}	 		\\
Mantle    							& 0.021\pm{0.019}	& 0.0059\pm{0.011}		&  0.0	 				& - 						& 0.027\pm{0.030}			\\ \hline
$\sum_\mathrm{Crust,Mantle}$     	& 0.035^{+0.023}_{-0.022}	& 0.014^{+0.014}_{-0.013}  		&  0.0					& 0.006^{+0.001}_{-0.001} 	& 0.055^{+0.037}_{-0.035} 			\\
\end{tabular}
\caption{Contribution of geoneutrino luminosities $L$ in AGM2015 above the IBD threshold $E_{\nu}\geq 1.8$ MeV for $^{238}$U, $^{232}$Th, and $^{40}$K $\nuebar$ emitted by the Earth. The reactor-$\nuebar$ luminosity $\geq 1.8$ MeV is $0.6 \pm 0.1 \times 10^{23} \nuebar / s$ and 26 GW for the world's 435 reactor cores, which together output 870 GW$_\mathrm{th}$\cite{PRIS2014}. Uncertainties shown in this table derived from Huang {\it et al.} \cite{hua13}, Arevalo {\it et al.} \cite{are09}, Gando {\it et al.} \cite{kam13}, and Bellini {\it et al.} \cite{bor13}. }
\label{table:luminosities2}
\end{table*}


\section*{Uncertainty}

The underlying interior structure and composition of the Earth is, in some regards, still poorly understood. The concentration and distribution of radioisotopes, whose decay chains produce geoneutrino flux, dominate the uncertainties. Therefore modeling of the distribution, energy spectra, and total flux of geoneutrinos remains a challenging task on its own. A full description of the uncertainty in each element of the AGM flux maps is not available at this time, however, we have defined the uncertainties in the specific building blocks of the AGM in Tables \ref{table:reactor_elements} and \ref{table:geo_elements} as well as the systematic uncertainties present in the various geoneutrino luminosity categories in Tables \ref{table:luminosities} and \ref{table:luminosities2}. The uncertainties in Tables \ref{table:reactor_elements} and \ref{table:geo_elements} in particular can be used to create Monte Carlo instances of the AGM flux maps, which could be used to evaluate the variances in each element of the AGM flux map, as well as co-variances between map elements. Such a full-scale Monte Carlo covariance matrix is impractical, however, due to the large number of map elements, $\sim$300 Million, which would end up producing a 300 Million $\times$ 300 Million full-size covariance matrix.

In most parts of the AGM2015 map uncertainties are strongly correlated over space and energy. This is due to the fact that the greatest uncertainty lies in ingredients that affect all map elements nearly equally, such as elemental abundances in the Earth's crust and mantle. Other ingredients which might introduce more independent uncertainty, such as the volume or density of specific crust tiles, are much more likely to introduce minimal correlations, or only slight regional correlations. Since reactor-$\nuebar$ flux is generally better predicted than geo-$\nuebar$ flux, and since near a core the reactor-$\nuebar$ flux will dominate the overall $\nuebar$ flux, we would expect smaller fractional uncertainties near reactors than in other regions of the world. 

We attempt to apply systematic uncertainties to AGM2015 where appropriate, and to derive these uncertainties from previous work in the field where possible rather than reinvent the wheel. For our uncertainty models we turn to Huang {\it et al.} \cite{hua13} and Dye \cite{dye15}. Doing so allows us to apply systematic uncertainties to the various $\nuebar$ source categories in Tables \ref{table:luminosities} and \ref{table:luminosities2} while avoiding the significant computational burden of a full Monte Carlo analysis, as well as the questions that would arise afterward of how to describe the various levels of regional correlation in flux uncertainty across space and energy. 

\section*{Conclusion}

Electron antineutrino measurements have allowed for the direct assessment of 7-29TW \cite{dye15} power from U and Th along with constraining a geo-reactor $<$ 3.7 TW at the 95\% confidence level \cite{kam13}. Such measurements promise the fine-tuning of BSE abundances and the distribution of heat-producing elements within the crust and mantle. Several such models of the Earth's antineutrino flux \cite{mantovani_2004,eno05,fog05} existed before the observation of geoneutrinos; with several recent models being presented with the inclusion of geoneutrinos \cite{hua13,sra13,fio12}. All of the aforementioned models incorporate several geophysical models based on the crust and mantle from traditional geophysical measurements (seismology, chondritic meteorites, etc.) This effort, AGM2015, aims to consolidate all these models into a user-friendly interactive map, freely available to the general public and easily accessible to anyone with a simple web browser at \href{http://www.ultralytics.com/agm2015}{www.ultralytics.com/agm2015}.

Future work includes completing a more detailed uncertainty study using Monte Carlo methods. Such a study requires an accurate understanding of the uncertainty in each of the AGM elements listed in Tables \ref{table:reactor_elements} and \ref{table:geo_elements}, however, as well as the correlations that govern their interactions. Future geo-$\nuebar$ measurements, updated flavor oscillation parameters, advances in crust/mantle models, and the ongoing construction and decommissioning of nuclear reactors around the world necessitates a dynamic AGM map capable of changing with the times. For this reason we envision the release of periodic updates to the original AGM2015, which will be labeled accordingly by the year of their release (i.e. ``AGM2020").

\section*{Methods}

AGM2015 uses CRUST1.0 \cite{crust1.0} to model the Earth's crustal density and volume profile via eight stratified layers. Elemental abundances for U, Th and K, and isotopic abundances for $^{238}$U, $^{232}$Th and $^{40}$K for each layer were defined by Huang \textit{et al.} \cite{hua13}. These values were coupled to well known isotope half-lives and multiplicities to create $\nuebar$ luminosities emanating from each crust tile. A similar approach was taken with the Earth's mantle, with elemental abundances derived via estimates of geo-$\nuebar$ flux at KamLAND and Borexino\cite{lud13} and density profiles supplied via PREM\cite{PREM}.

Man-made reactors were modeled via the IAEA PRIS\cite{PRIS2014} database, with reactor-$\nuebar$ luminosities found to scale as $1.83 \times 10^{20}$ $\nuebar/\mathrm{s}/\mathrm{GW_{th}}$. Reactor-$\nuebar$ spectra were modeled as exponential falloffs from empirical data\cite{ber10}, while $^{238}$U, $^{232}$Th and $^{40}$K spectra were modeled based on the work of Sanshiro Enomoto\cite{Enomoto_Sanshiro}.

Luminosities from each point-source were converted to fluxes at each map location via the $P_{e \rightarrow e}$ survival probability shown in Equation \ref{osceqn1}, and a full understanding of the source spectra of each point-source enabled a complete reconstruction of the observed energy spectra at each map location. ``Smart" integration was applied where necessary to more accurately portray crust and mantle tiles as volume-sources rather than point-sources. All modeling and visualization was done with MATLAB\cite{matlab}. Google Maps and Google Earth multi-resolution raster pyramids created with MapTiler\cite{maptiler}. All online content available at \href{http://www.ultralytics.com/agm2015}{www.ultralytics.com/agm2015}.

\section*{Acknowledgements}
We would like to thank S. Enomoto \& M. Sakai for discussions at various phases of this project. This work is supported in part by a National Science Foundation grants EAR 1068097 and EAR 1067983, the U.S. Department of Energy, and the National Geospatial-Intelligence Agency.

\section*{Author Contributions}
S.M.U. suggested this study. G.R.J. conducted all simulations and incorporated models written by S.D., W.F.M., and J.G.L. Mapping formats and standards were developed by G.R.J. and S.M.U. All authors contributed to writing the paper. 

\section*{Additional Information}
Competing financial interests: The authors declare no competing financial interests.

\end{document}